\documentclass{aac}

\usepackage[numbers]{natbib}
\usepackage{rotating}
\usepackage{graphicx}

\begin{document}

\title{Diagnostic Techniques for Particle-in-Cell Simulations of Laser-produced Gamma-rays in the Strong-field QED Regime}

\author[aff1,aff2]{A. H. Younis\corref{cor1}}
\author[aff2]{A. Davidson}
\author[aff2]{B. Hafizi}
\author[aff2]{D. F. Gordon}

\affil[aff1]{Department of Physics and Astronomy, University of Rochester, Rochester, New York 14627, USA}
\affil[aff2]{Plasma Physics Division, Naval Research Laboratory, Washington D.C. 20375, USA}
\corresp[cor1]{ayounis@ur.rochester.edu}

\maketitle

\begin{abstract}
The emission of multi-MeV ($\gamma$-ray) photons from the interaction of a high-powered laser pulse with a dense plasma target is studied using particle-in-cell simulations. A new set of diagnostic techniques is presented and applied to analyze the intense field and ultra-relativistic electrons. Such methods elucidate the dominant processes responsible for efficient laser-to-$\gamma$ energy conversion, which include nonlinear Compton scattering and magneto-bremsstrahlung radiation, and provide a clear picture of the interaction on a microscopic level. We identify regions in the plasma target of high photon energy-density and obtain an energy conversion efficiency as high as 30\%. The essential characteristics of the interaction are validated with full-3D simulations.
\end{abstract}

\section{INTRODUCTION}
The study of laser-plasma interactions in the ultra-relativistic regime is an active frontier of research with close ties to nonlinear optics and high-energy particle physics \cite{Mourou-2006}. In the strong-field limit, approaching the Schwinger critical field in the particle rest frame, quantum electrodynamical effects govern the interaction and novel processes emerge as the vacuum itself begins to respond nonlinearly \cite{Mourou-2006, Bulanov-2015}. Next-generation laser systems like the Extreme Light Infrastructure (ELI) facilities in Europe \cite{Tanaka-2020} will operate in the intensity range of $10^{23}$--$10^{24}\textrm{ W/cm}^2$ and enable production of QED-dominated plasmas in a controlled laboratory environment. Therefore, processes that are suppressed exponentially for field strengths well-below the Schwinger limit, like Breit-Wheeler pair creation and nonlinear Compton scattering, will become accessible for experimental study due to the high particle energies that can be obtained \cite{Gu-2019}.
\par
In recent years, the particle-in-cell (PIC) approach to numerical plasma modeling \cite{Dawson-1983, Arber-2015} has been extended to account for quantum effects. The implementation of QED-PIC routines is reviewed extensively in, e.g., \cite{Sokolov-2011, Ridgers-2014, Gonoskov-2015} and references contained therein. They have also been successful at providing insights into processes like energetic photon emission \cite{Nakamura-2012, Ridgers-2012}, pair creation \cite{Ridgers-2012, Vranic-2018}, and quantum radiation-reaction \cite{Vranic-2014, Vranic-2016}. In this paper, we introduce a new set of PIC diagnostic methods to study the $\gamma$-ray emission mechanism in particular, paying close attention to nonlinear Compton scattering and magneto-bremsstrahlung radiation. One of the primary results, consistent with other numerical studies \cite{Lezhnin-2018, Arefiev-2020}, is that an unprecedented laser-to-$\gamma$ energy conversion efficiency on the order of 10\% is attainable with field strengths that are just within reach of modern experimental facilities. Our latest results highlight the precise location of bulk $\gamma$-ray production within the plasma target as well as the interaction between the ultra-relativistic particles and the scattered and quasi-static field components. The techniques are applied in the post-processing stage, and so they are usable with any PIC code capable of exporting reduced-domain pseudo-particle and field data. Our results were obtained with the code EPOCH \cite{Arber-2015} and verified with OSIRIS \cite{osiris-paper} for consistency and extendability.
\par
In what follows, we will briefly summarize the theory and numerical implementation of photon-emitting processes. Then we will discuss the model plasma target and simulation parameters, followed by an outline of each diagnostic method, the resulting data, and its significance. We will conclude with general remarks and discuss future directions.

\subsection{Photon Emission: Theory and Numerical Implementation}
The primary mechanisms responsible for high-energy photon emission are nonlinear Compton scattering and magneto-bremsstrahlung radiation. The high photon flux from ultra-intense laser fields, characterized by the normalized amplitude $a_0=e\sqrt{-A_\mu A^\mu}/mc^2$ being significantly greater than unity, leads to nonlinear Compton scattering in which an electron absorbs $n\gg1$ laser photons $\omega_0$ before emitting a photon $\omega_\gamma$ of its own \cite{DiPiazza-2011}. (Standard notation is adopted throughout; $e$ and $m$ are the electron charge magnitude and mass, $c$ is the speed of light, and $A^\mu$ is the field 4-vector potential.) In cases where the emitted photon energy is comparable to the electron rest mass, the electron also experiences a non-negligible recoil which must be taken into account. For dense plasmas, the current produced by the collective motion of ultra-relativistic electrons can drive a quasi-static magnetic field strong enough to induce magneto-bremsstrahlung radiation. The dynamical behavior of electrons in such a scenario is referred to as a forward sliding-swing acceleration, and it arises in situations where the plasma current is significantly greater than the Alfv\'{e}n limit $I_A=\beta\gamma mc^3/e$, where $\beta=|\vec{v}\,|/c$ is the normalized velocity, and $\gamma=1/\sqrt{1-\beta^2}$ is the Lorentz factor. The quasi-static field strength is reduced from the MT-scale oscillating field by an order of magnitude, and the resulting transverse confinement of electrons leads to highly-collimated $\gamma$-ray beams \cite{Arefiev-2020}.
\par
In terms of numerical implementation, analytic expressions for the $S$-matrix elements are used to derive an emission probability rate and the stochastic nature of the process is captured by a Monte Carlo algorithm \cite{Ridgers-2014}. The overall emission probability, accounting for all processes through a generic field term, depends on the parameters:
\begin{equation}
\eta = \frac{e\hbar}{m^3c^4}|F_{\mu\nu}p^\nu|\quad\textrm{and}\quad \chi = \frac{e\hbar^2}{m^3c^4}|F_{\mu\nu}k^\nu|,
\label{eqn:QuantumParams}
\end{equation}
which characterize the strength of nonlinear QED effects for electrons and photons, respectively. Here, $F_{\mu\nu}=\partial_\mu A_\nu - \partial_\nu A_\mu$ is the electromagnetic field tensor, and $p^\nu$ ($\hbar k^\nu$) is the 4-momentum of the electron (photon). Alternatively, $\eta = E_\textsc{rf}/E_\textsc{s}$, where $E_\textsc{rf}\sim\gamma E$ is the field magnitude in the electron rest frame, and $E_\textsc{s}=m^2c^3/\hbar e\simeq1.32\times10^{18}\textrm{ V/m}$ is the Schwinger critical field strength. These parameters serve as a measure of energy in the joint particle-field system, though it must be remembered that the 4-vector magnitudes result in an angular dependence between the field and momentum components. The quantum parameters are maximized if the particle momentum is anti-parallel to the field wavevector $\vec{k}$, and minimized if they are parallel: $\eta\,_\textsc{em} = \gamma(E \pm \beta B)/E_\textsc{s}$, and similarly for $\chi$. Typical values of $\eta$ in our simulations range from $10^{-1}$ and below.
\par
\begin{figure}[b]
\centerline{\includegraphics[height=0.185\textheight]{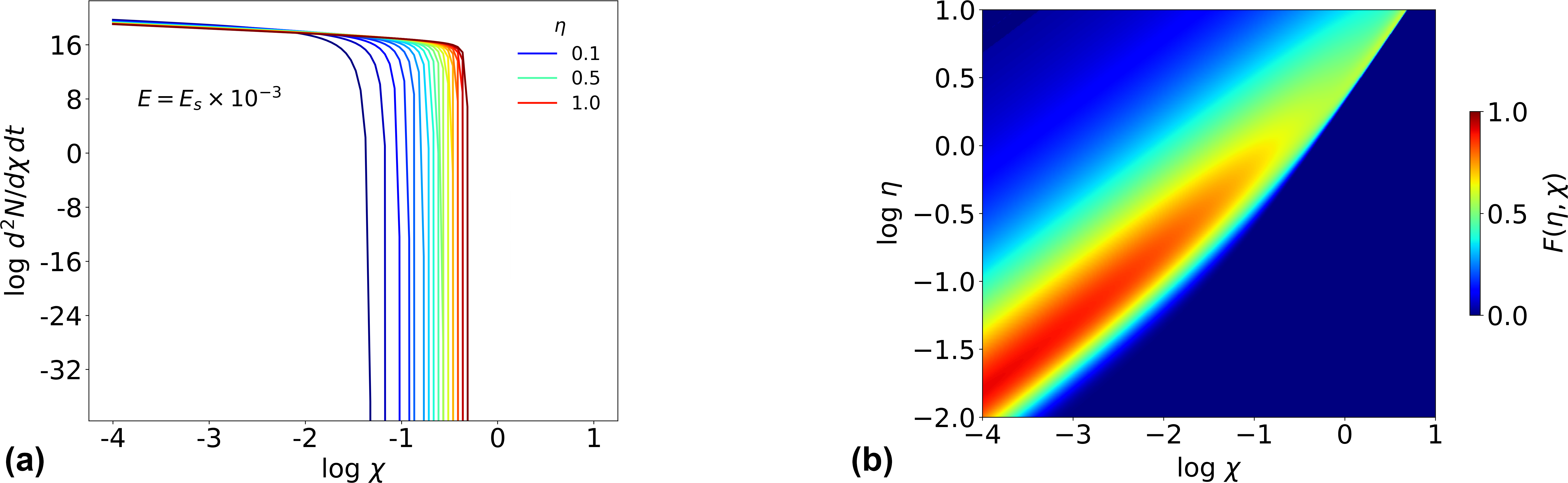}}
\caption{(a) Differential photon emission rate. (b) Synchrotron function vs. electron $\eta$ and photon $\chi$ quantum parameters.}
\label{fig:SynchFunc}
\end{figure}

In our case, we are interested in $\eta$ as a measure of the photon emission probability. By recording pseudo-particle momentum data and interpolating the field to their positions, we can determine which components result in high-$\eta$ electrons as the interaction unfolds. Given $\eta$, the spin-averaged emission rate of photons about a $d\chi$ interval is:
\begin{equation}
\frac{d^2N}{d\chi\,dt} = \sqrt{3}\frac{mc^2}{\hbar}\alpha b\frac{F(\eta,\chi)}{\chi},\quad F(\eta,\chi) = \frac{4\chi^2}{\eta^2}yK_{2/3}(y) + \Big(1-\frac{2\chi}{\eta}\Big)\,y\int_{y}^{\infty}dt\,K_{5/3}(t),
\label{eqn:DiffEmRate}
\end{equation}
where $\alpha=e^2/\hbar c$ is the fine-structure constant, $b=E/E_\textsc{s}$ is the normalized field amplitude, and $F(\eta,\chi)$ is the synchrotron function which depends on $y=4\chi/[3\eta(\eta-2\chi)]$, and on modified Bessel functions of the second kind $K_n(x)$ \cite{Ridgers-2014, DiPiazza-2011, LL-QED}. By Eq.~(\ref{eqn:DiffEmRate}), the most probable emitted-photon quantum parameter increases with $\eta$, though the emission rate varies inversely with $\chi$ so that high-energy photons are relatively rare (see Fig.~\ref{fig:SynchFunc}).

\clearpage

\section{RESULTS}
\subsection{Simulation Setup}
\begin{figure}[h]
\centerline{\includegraphics[width=0.3\textwidth]{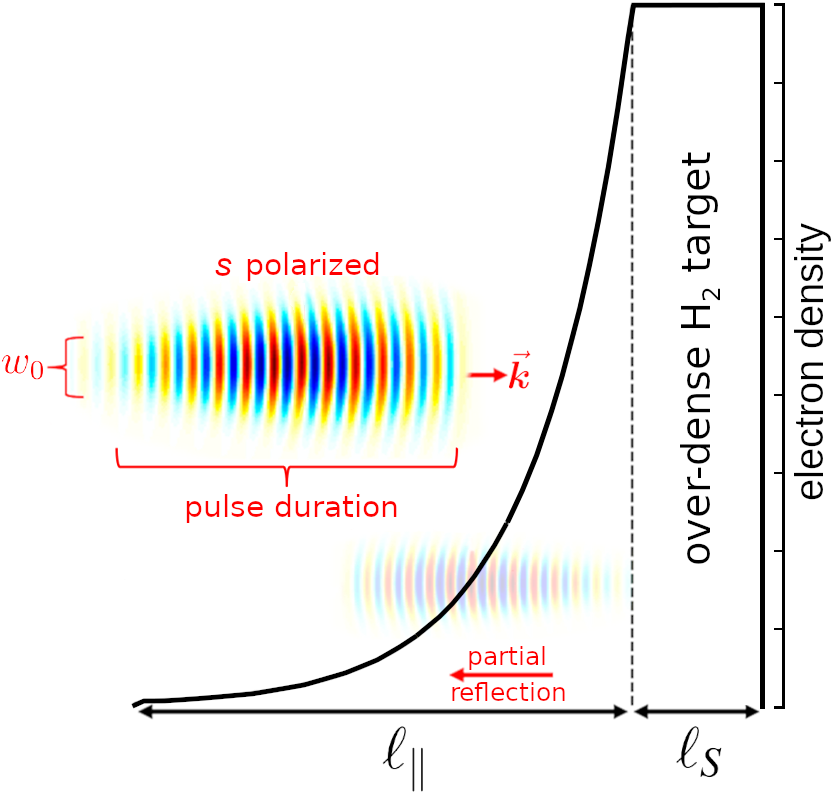}}
\caption{Illustration of the plasma profile used to study $\gamma$-ray emission.}
\label{fig:SimulationSetup}
\end{figure}

The setup used to study $\gamma$-ray emission is shown in Fig.~\ref{fig:SimulationSetup}. A 30 fs, 20 PW pulse ($I\simeq 2\times10^{23}\textrm{ W/cm}^2$) polarized in the transverse direction is incident on a plasma target whose density increases exponentially to $20n_{cr}$ over a distance of $\ell_\parallel=40\textrm{ }\mu\textrm{m}$. Here, $n_{cr}=m\omega_0^2/4\pi e^2$ is the critical density associated with the laser frequency $\omega_0$. In this case, the vacuum wavelength is $\lambda=1\textrm{ }\mu\textrm{m}$, the Gaussian beam waist is $w=5\lambda$, and $n_{cr}\simeq1.12\times10^{21}\,\textrm{ cm}^{-3}$. Moreover, the over-dense target slab is $\ell_S=10\textrm{ }\mu\textrm{m}$ thick. These values are optimal for a high laser-to-$\gamma$ energy conversion efficiency (30\%) and they are based on the multi-parametric studies of Lezhnin \textit{et al.} \cite{Lezhnin-2018}. While our diagnostics are applicable to 2D simulations, we also performed full-3D runs on a $90\lambda\textrm{ (L) }\times20\lambda\textrm{ (W) }\times20\lambda\textrm{ (H)}$ grid utilizing 12 $\textrm{nodes}/\lambda$ and 6 pseudo-particles per cell. The threshold energy required for photon emission was set to $mc^2/10\simeq51.1\textrm{ keV}$. This is a purely computational limit used to prevent the simulation from overflowing with pseudo-particles and encountering memory issues. We also chose only to record where photons were emitted, neglecting their propagation and any subsequent interactions. Of course, if pair-creation or other QED processes were of interest this would not be valid.
\vspace{2em}
\vfill
\subsection{Field Analysis}
\begin{figure}[t]
\hspace{-2em}\centerline{\includegraphics[height=0.2\textheight]{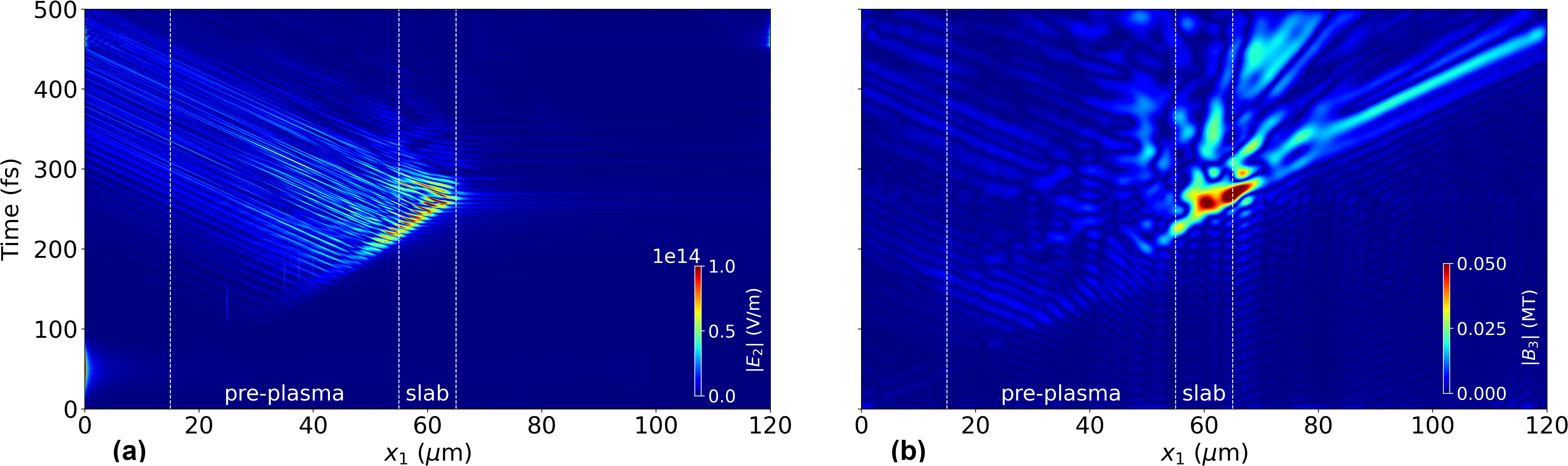}}
\caption{Plots in $t$-$x_1$ space of (a) the reflected transverse electric field $E_2$ and (b) the quasi-static azimuthal magnetic field $B_3$. The spatial profile of the plasma target is overlaid in white. The quasi-static field was obtained by inverse-transforming the region: $-\omega_0/4\leq\omega\leq\omega_0/4$ ($0\pm75\textrm{ THz}$) and $-|\mathbf{k}|/4\leq k_1\leq |\mathbf{k}|/4$ ($0\pm\pi/2\textrm{ }\mu\textrm{m}^{-1}$).}
\label{fig:FilteredLineoutData}
\end{figure}

The first technique we developed to study photon emission involves isolating the reflected and quasi-static components of the combined laser-plasma field. We recorded 1D on-axis lineouts of each spatial component at a rate of 50 attoseconds (20 PHz). By performing a 2D Fourier transform on the resulting coordinate-time data $(t,x_1)\to(\omega,k_1)$ and filtering regions in frequency-space where $\omega/k_1>0$, we isolated the reflected field [Fig.~\ref{fig:FilteredLineoutData}(a)]. (All throughout, subscript 1 denotes the longitudinal pulse-propagation axis, subscript 2 the transverse axis, and subscript 3 the axis orthogonal to 1 and 2.) A similar filtering method about the zero-frequency signal was employed to capture the slowly-varying nature of $B_3$, the orthogonal magnetic field [Fig.~\ref{fig:FilteredLineoutData}(b)].
\par
From these diagrams, we see that there is a small amount of reflection occurring within the pre-plasma which increases as the pulse approaches the dense target slab. This is expected, as the plasma is partially transparent to such an ultra-relativistic pulse. The reflected field amplitude is an order of magnitude smaller than the incident component, and it is strongly attenuated within the slab beyond $t\simeq 300$ fs. Moreover, it has a total lifetime of about 100 fs. As we will discuss below, the attenuation aligns spatially and temporally with $\gamma$-ray production in the plasma, i.e., energy from the reflected field is depleted to generate $\gamma$-rays. In light of this and the theory discussed in the previous section, we believe that optimizing the field reflection of the interior region of the target will increase the overall conversion efficiency and peak $\gamma$ energy. This may motivate future multi-parametric studies. Lastly, the quasi-static magnetic field has a maximum of approximately 0.1 MT which is capable of inducing magneto-bremsstrahlung radiation, though the dominant mechanism here is nonlinear Compton scattering due to the pulse intensity and target geometry under consideration \cite{Lezhnin-2018}. Transverse electron confinement is less pronounced for a slab geometry than, e.g., a plasma waveguide structure \cite{Arefiev-2020}, which reduces the efficiency of the magneto-bremsstrahlung process.

\clearpage

\begin{figure}[h!]
\hspace{-2em}\centerline{\includegraphics[height=0.2\textheight]{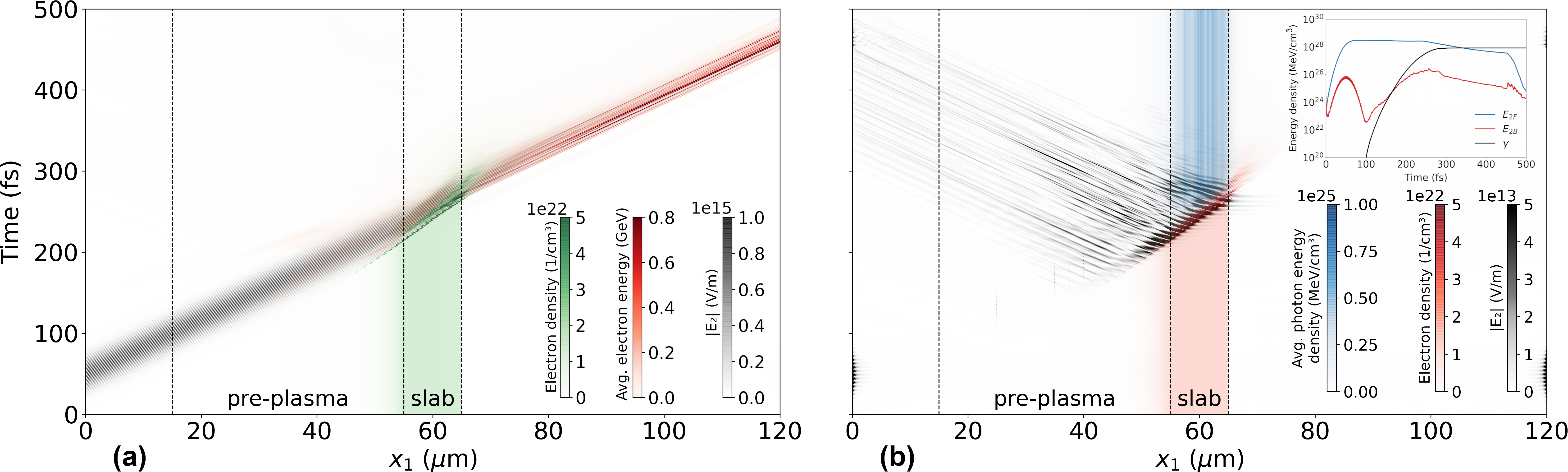}}
\caption{Overlay diagrams in $t$-$x_1$ space of various quantities, including the full and reflected transverse electric field, the average electron energy, and the photon energy-density. (a) The pulse bores through the dense slab and accelerates electrons to GeV-scale energies. (b) Interaction between the reflected field and energetic electrons leads to $\gamma$-ray emission via nonlinear Compton scattering. (b) (inset) Energy vs.~time of the forward/reflected field and $\gamma$ rays. Note: The exponentially-decaying pre-plasma is not visible on this scale, nor is any photon emission occurring within the pre-plasma or beyond the slab.}
\label{fig:OverlayDiagram}
\end{figure}

\begin{figure}[h!]
\hspace{-2em}\centerline{\includegraphics[height=0.2\textheight]{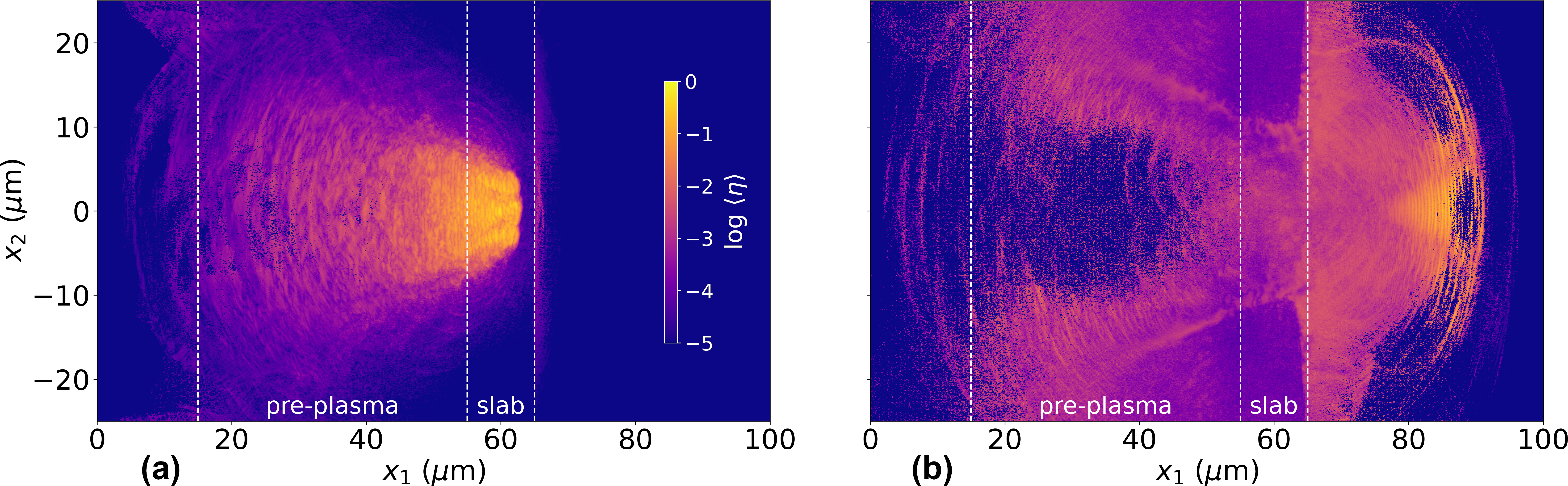}}
\caption{Average electron quantum parameter at (a) the instant of maximum $\gamma$-ray emission, $t=250$ fs and (b) $100$ fs later.}
\label{fig:QParamAvg}
\end{figure}

\subsection{Overlay Diagrams}
Extending the lineout diagnostic, particle data including average energy and number density was recorded. Overlaying these quantities in $t$-$x_1$ space provides valuable and unique insights into this process. In Fig.~\ref{fig:OverlayDiagram}(a), lineouts of the full transverse electric field, electron density, and average electron energy are overlaid together. This diagram illustrates how the pulse expels electrons from the target completely, accelerating them to sub-GeV energies. Within the target slab, the transverse field distorts and decreases in strength owing to the strong reflection noted above. In Fig.~\ref{fig:OverlayDiagram}(b), the average photon energy-density is plotted together with the electron density. It must be remembered that our simulations were performed with photon propagation turned off, in order to allow us to observe the precise location of $\gamma$-ray emission. The highest energy photons are emitted within the target slab, though $\gamma$-ray radiation also occurs beyond it as the relativistic electrons continue interacting with the field. Lastly, Fig.~\ref{fig:OverlayDiagram}(b) also shows the reflected transverse field from Fig.~\ref{fig:FilteredLineoutData}(a). Here, we see the spatiotemporal alignment between the two key events: rapid decay of the reflected field and emission of $\gamma$-rays.

\subsection{Quantum Parameter Tracking}
\begin{figure}[h]
\centerline{\includegraphics[width=0.675\textwidth]{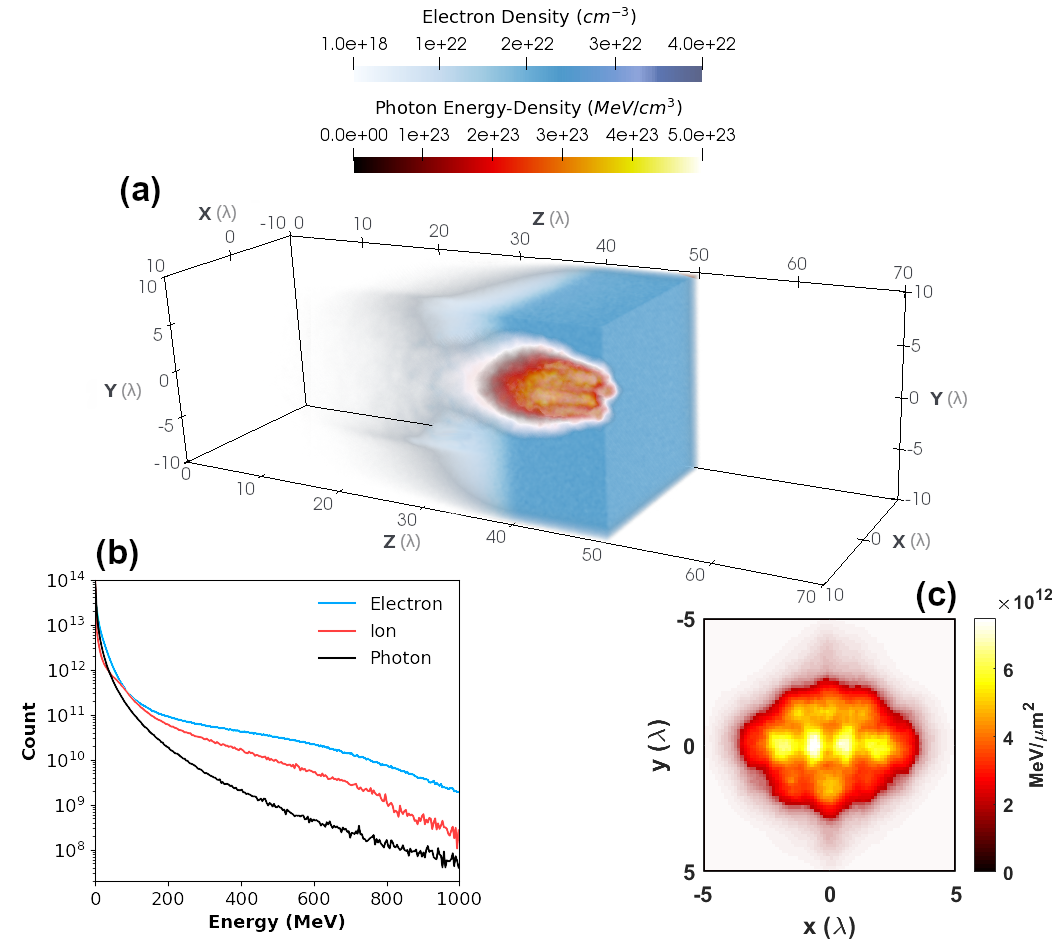}}
\caption{(a) Isometric view of the plasma target and $\gamma$-ray energy density at the instant of maximum radiated power. (b) Particle energy spectrum. (c) Total $\gamma$ fluence in the direction of pulse propagation.}
\label{fig:3DPlot}
\end{figure}

Another diagnostic we developed evaluates the electron quantum parameter [see Eq.~(\ref{eqn:QuantumParams})] for all $\sim$20 million pseudo-particles in the simulation (Fig.~\ref{fig:QParamAvg}). We record the pseudo-particle Lorentz factor and momentum components for $p^\nu$, and bicubic interpolation \cite{NumericalRecipes} is used to obtain the tensor components $F_{\mu\nu}$ at each pseudo-particle position. We also tracked individual electron trajectories over time. It was found that as an electron propagates through the dense target slab at $t\simeq250$ fs, it obtains a high quantum parameter on the order of $10^{-1}$ which aligns temporally with the instant of peak photon emission. (This value may not seem significant, but if one recalls the equivalent definition $\eta=E_\textsc{rf}/E_\textsc{s}$, where $E_\textsc{rf}$ is the field in the particle rest frame, then one sees that we are dealing with non-negligible fractions of the Schwinger field, i.e. the onset of nonlinear QED.) Upcoming work with our particle-tracking diagnostic involves determining the average radiated power for a group of closely-initialized electrons, and comparing it with other groups initialized in different regions. However, the dynamical behavior is highly nonlinear, and so consistency in radiated power between electron pseudo-particles initialized relatively close together may not be guaranteed. Lastly, Fig. \ref{fig:3DPlot} shows a three-dimensional view of the $\gamma$-flare, from which it is apparent that there are longitudinal columns where $\gamma$ production is the most efficient. This is due to a combination of relativistic self-focusing of the pulse and electron expulsion away from the central axis. With full photon dynamics included, the $\gamma$ flash propagates out in a spherically-symmetric manner but the flash-front has the highest energy photons.
\vfill

\clearpage

\section{CONCLUSION}
The interaction between an ultra-relativistic field and an over-critical plasma target presents the opportunity to study nonlinear QED effects, such as intense $\gamma$-ray emission. Using particle-in-cell simulations, we have developed new techniques to analyze this process and discovered that the scattered field plays a central role in maximizing the photon emission probability. It then becomes important to understand how the energy transfer between the forward-propagating field, whose primary role includes accelerating particles to the required energy levels, and the back-scattered field, which induces photon emission, can be optimized. This may motivate future multi-parametric studies with hopes that any insights gained about the plasma target properties will guide experimental campaigns. Our main technique can also be used to isolate the quasi-static nature of any field component, which is important for photon emission studies based on the magneto-bremsstrahlung process. Unfortunately, it is limited in that information can only be extracted from narrow sections of the domain. In the near future, we may explore ways of extending this diagnostic. Finally, we observed longitudinal columns where $\gamma$-rays are generated without any pre-manufactured electron-confining structures in the target geometry. These energetic photons can be used for other studies in fundamental physics, such as relativistic photo-ionization and Breit-Wheeler pair creation.

\section{ACKNOWLEDGMENTS}
This work was supported by the US DOE Office of Science under Interagency Agreement number 89243018SSC000006 and the Directed Energy Society's Directed Energy Summer Internship program. This research was conducted while A. D. held an NRC Research Associateship award at the Naval Research Laboratory. Simulations were performed on the National Energy Research Scientific Computing (NERSC) center's Cori computing cluster.

\nocite{*}
\bibliographystyle{aac}%
\bibliography{References}%

\begin{thebibliography}{21}%
\makeatletter
\providecommand \@ifxundefined [1]{%
 \@ifx{#1\undefined}
}%
\providecommand \@ifnum [1]{%
 \ifnum #1\expandafter \@firstoftwo
 \else \expandafter \@secondoftwo
 \fi
}%
\providecommand \@ifx [1]{%
 \ifx #1\expandafter \@firstoftwo
 \else \expandafter \@secondoftwo
 \fi
}%
\providecommand \natexlab [1]{#1}%
\providecommand \enquote  [1]{``#1''}%
\providecommand \bibnamefont  [1]{#1}%
\providecommand \bibfnamefont [1]{#1}%
\providecommand \citenamefont [1]{#1}%
\providecommand \href@noop [0]{\@secondoftwo}%
\providecommand \href [0]{\begingroup \@sanitize@url \@href}%
\providecommand \@href[1]{\@@startlink{#1}\@@href}%
\providecommand \@@href[1]{\endgroup#1\@@endlink}%
\providecommand \@sanitize@url [0]{\catcode `\$12\catcode `\&12\catcode
  `\#12\catcode `\^12\catcode `\_12\catcode `\%12\relax}%
\providecommand \@@startlink[1]{}%
\providecommand \@@endlink[0]{}%
\providecommand \url  [0]{\begingroup\@sanitize@url \@url }%
\providecommand \@url [1]{\endgroup\@href {#1}{\urlprefix }}%
\providecommand \urlprefix  [0]{URL }%
\providecommand \Eprint [0]{\href }%
\providecommand \doibase [0]{http://dx.doi.org/}%
\providecommand \selectlanguage [0]{\@gobble}%
\providecommand \bibinfo  [0]{\@secondoftwo}%
\providecommand \bibfield  [0]{\@secondoftwo}%
\providecommand \translation [1]{[#1]}%
\providecommand \BibitemOpen [0]{}%
\providecommand \bibitemStop [0]{}%
\providecommand \bibitemNoStop [0]{.\EOS\space}%
\providecommand \EOS [0]{\spacefactor3000\relax}%
\providecommand \BibitemShut  [1]{\csname bibitem#1\endcsname}%
\let\auto@bib@innerbib\@empty
\bibitem [{\citenamefont {Mourou}, \citenamefont {Tajima},\ and\ \citenamefont
  {Bulanov}(2006)}]{Mourou-2006}%
  \BibitemOpen
  \bibfield  {author} {\bibinfo {author} {\bibfnamefont {G.~A.}\ \bibnamefont
  {Mourou}}, \bibinfo {author} {\bibfnamefont {T.}~\bibnamefont {Tajima}}, \
  and\ \bibinfo {author} {\bibfnamefont {S.~V.}\ \bibnamefont {Bulanov}},\
  }\href@noop {} {\bibfield  {journal} {\bibinfo  {journal} {Rev. Mod. Phys.}\
  }\textbf {\bibinfo {volume} {78}},\ p.\ \bibinfo {pages} {309} (\bibinfo
  {year} {2006})}\BibitemShut {NoStop}%
\bibitem [{\citenamefont {Bulanov}\ \emph {et~al.}(2015)\citenamefont {Bulanov}
  \emph {et~al.}}]{Bulanov-2015}%
  \BibitemOpen
  \bibfield  {author} {\bibinfo {author} {\bibfnamefont {S.~V.}\ \bibnamefont
  {Bulanov}} \emph {et~al.},\ }\href@noop {} {\bibfield  {journal} {\bibinfo
  {journal} {Plasma Physics Reports}\ }\textbf {\bibinfo {volume} {41}},\
  \unskip\ \bibinfo {pages} {1--51} (\bibinfo {year} {2015})}\BibitemShut
  {NoStop}%
\bibitem [{\citenamefont {Tanaka}\ \emph {et~al.}(2020)\citenamefont {Tanaka}
  \emph {et~al.}}]{Tanaka-2020}%
  \BibitemOpen
  \bibfield  {author} {\bibinfo {author} {\bibfnamefont {K.~A.}\ \bibnamefont
  {Tanaka}} \emph {et~al.},\ }\href@noop {} {\bibfield  {journal} {\bibinfo
  {journal} {Matter Radiat. Extremes}\ }\textbf {\bibinfo {volume} {5}},\ p.\
  \bibinfo {pages} {024402} (\bibinfo {year} {2020})}\BibitemShut {NoStop}%
\bibitem [{\citenamefont {Gu.}\ \emph {et~al.}(2019)\citenamefont {Gu.},
  \citenamefont {Jirka}, \citenamefont {Klimo},\ and\ \citenamefont
  {Weber}}]{Gu-2019}%
  \BibitemOpen
  \bibfield  {author} {\bibinfo {author} {\bibfnamefont {Y.-J.}\ \bibnamefont
  {Gu.}}, \bibinfo {author} {\bibfnamefont {M.}~\bibnamefont {Jirka}}, \bibinfo
  {author} {\bibfnamefont {O.}~\bibnamefont {Klimo}}, \ and\ \bibinfo {author}
  {\bibfnamefont {S.}~\bibnamefont {Weber}},\ }\href@noop {} {\bibfield
  {journal} {\bibinfo  {journal} {Matter Radiat. Extremes}\ }\textbf {\bibinfo
  {volume} {4}},\ p.\ \bibinfo {pages} {064403} (\bibinfo {year}
  {2019})}\BibitemShut {NoStop}%
\bibitem [{\citenamefont {Dawson}(1983)}]{Dawson-1983}%
  \BibitemOpen
  \bibfield  {author} {\bibinfo {author} {\bibfnamefont {J.~M.}\ \bibnamefont
  {Dawson}},\ }\href@noop {} {\bibfield  {journal} {\bibinfo  {journal} {Rev.
  Mod. Phys.}\ }\textbf {\bibinfo {volume} {55}},\ p.\ \bibinfo {pages} {403}
  (\bibinfo {year} {1983})}\BibitemShut {NoStop}%
\bibitem [{\citenamefont {Arber}\ \emph {et~al.}(2015)\citenamefont {Arber}
  \emph {et~al.}}]{Arber-2015}%
  \BibitemOpen
  \bibfield  {author} {\bibinfo {author} {\bibfnamefont {T.~D.}\ \bibnamefont
  {Arber}} \emph {et~al.},\ }\href@noop {} {\bibfield  {journal} {\bibinfo
  {journal} {Plasma Phys. Control. Fusion}\ }\textbf {\bibinfo {volume} {57}},\
  p.\ \bibinfo {pages} {113001} (\bibinfo {year} {2015})}\BibitemShut {NoStop}%
\bibitem [{\citenamefont {Sokolov}, \citenamefont {Naumova},\ and\
  \citenamefont {Nees}(2011)}]{Sokolov-2011}%
  \BibitemOpen
  \bibfield  {author} {\bibinfo {author} {\bibfnamefont {I.~V.}\ \bibnamefont
  {Sokolov}}, \bibinfo {author} {\bibfnamefont {N.~M.}\ \bibnamefont
  {Naumova}}, \ and\ \bibinfo {author} {\bibfnamefont {J.~A.}\ \bibnamefont
  {Nees}},\ }\href@noop {} {\bibfield  {journal} {\bibinfo  {journal} {Phys.
  Plasmas}\ }\textbf {\bibinfo {volume} {18}},\ p.\ \bibinfo {pages} {093109}
  (\bibinfo {year} {2011})}\BibitemShut {NoStop}%
\bibitem [{\citenamefont {Ridgers}\ \emph {et~al.}(2014)\citenamefont {Ridgers}
  \emph {et~al.}}]{Ridgers-2014}%
  \BibitemOpen
  \bibfield  {author} {\bibinfo {author} {\bibfnamefont {C.~P.}\ \bibnamefont
  {Ridgers}} \emph {et~al.},\ }\href@noop {} {\bibfield  {journal} {\bibinfo
  {journal} {J. Comp. Phys.}\ }\textbf {\bibinfo {volume} {260}},\ p.\ \bibinfo
  {pages} {273} (\bibinfo {year} {2014})}\BibitemShut {NoStop}%
\bibitem [{\citenamefont {Gonoskov}\ \emph {et~al.}(2015)\citenamefont
  {Gonoskov} \emph {et~al.}}]{Gonoskov-2015}%
  \BibitemOpen
  \bibfield  {author} {\bibinfo {author} {\bibfnamefont {A.}~\bibnamefont
  {Gonoskov}} \emph {et~al.},\ }\href@noop {} {\bibfield  {journal} {\bibinfo
  {journal} {Phys. Rev. E}\ }\textbf {\bibinfo {volume} {92}},\ p.\ \bibinfo
  {pages} {023305} (\bibinfo {year} {2015})}\BibitemShut {NoStop}%
\bibitem [{\citenamefont {Nakamura}\ \emph {et~al.}(2012)\citenamefont
  {Nakamura} \emph {et~al.}}]{Nakamura-2012}%
  \BibitemOpen
  \bibfield  {author} {\bibinfo {author} {\bibfnamefont {T.}~\bibnamefont
  {Nakamura}} \emph {et~al.},\ }\href@noop {} {\bibfield  {journal} {\bibinfo
  {journal} {Phys. Rev. Lett.}\ }\textbf {\bibinfo {volume} {108}},\ p.\
  \bibinfo {pages} {195001} (\bibinfo {year} {2012})}\BibitemShut {NoStop}%
\bibitem [{\citenamefont {Ridgers}\ \emph {et~al.}(2012)\citenamefont {Ridgers}
  \emph {et~al.}}]{Ridgers-2012}%
  \BibitemOpen
  \bibfield  {author} {\bibinfo {author} {\bibfnamefont {C.~P.}\ \bibnamefont
  {Ridgers}} \emph {et~al.},\ }\href@noop {} {\bibfield  {journal} {\bibinfo
  {journal} {Phys. Rev. Lett.}\ }\textbf {\bibinfo {volume} {108}},\ p.\
  \bibinfo {pages} {165006} (\bibinfo {year} {2012})}\BibitemShut {NoStop}%
\bibitem [{\citenamefont {Vranic}\ \emph {et~al.}(2018)\citenamefont {Vranic},
  \citenamefont {Klimo}, \citenamefont {Korn},\ and\ \citenamefont
  {Weber}}]{Vranic-2018}%
  \BibitemOpen
  \bibfield  {author} {\bibinfo {author} {\bibfnamefont {M.}~\bibnamefont
  {Vranic}}, \bibinfo {author} {\bibfnamefont {O.}~\bibnamefont {Klimo}},
  \bibinfo {author} {\bibfnamefont {G.}~\bibnamefont {Korn}}, \ and\ \bibinfo
  {author} {\bibfnamefont {S.}~\bibnamefont {Weber}},\ }\href@noop {}
  {\bibfield  {journal} {\bibinfo  {journal} {Scientific Reports}\ }\textbf
  {\bibinfo {volume} {8}},\ p.\ \bibinfo {pages} {4702} (\bibinfo {year}
  {2018})}\BibitemShut {NoStop}%
\bibitem [{\citenamefont {Vranic}\ \emph {et~al.}(2014)\citenamefont {Vranic}
  \emph {et~al.}}]{Vranic-2014}%
  \BibitemOpen
  \bibfield  {author} {\bibinfo {author} {\bibfnamefont {M.}~\bibnamefont
  {Vranic}} \emph {et~al.},\ }\href@noop {} {\bibfield  {journal} {\bibinfo
  {journal} {Phys. Rev. Lett.}\ }\textbf {\bibinfo {volume} {113}},\ p.\
  \bibinfo {pages} {134801} (\bibinfo {year} {2014})}\BibitemShut {NoStop}%
\bibitem [{\citenamefont {Vranic}\ \emph {et~al.}(2016)\citenamefont {Vranic},
  \citenamefont {Grismayer}, \citenamefont {Fonseca},\ and\ \citenamefont
  {Silva}}]{Vranic-2016}%
  \BibitemOpen
  \bibfield  {author} {\bibinfo {author} {\bibfnamefont {M.}~\bibnamefont
  {Vranic}}, \bibinfo {author} {\bibfnamefont {T.}~\bibnamefont {Grismayer}},
  \bibinfo {author} {\bibfnamefont {R.}~\bibnamefont {Fonseca}}, \ and\
  \bibinfo {author} {\bibfnamefont {L.}~\bibnamefont {Silva}},\ }\href@noop {}
  {\bibfield  {journal} {\bibinfo  {journal} {New J. Phys.}\ }\textbf {\bibinfo
  {volume} {18}},\ p.\ \bibinfo {pages} {073035} (\bibinfo {year}
  {2016})}\BibitemShut {NoStop}%
\bibitem [{\citenamefont {Lezhnin}\ \emph {et~al.}(2018)\citenamefont
  {Lezhnin}, \citenamefont {Sasorov}, \citenamefont {Korn},\ and\ \citenamefont
  {Bulanov}}]{Lezhnin-2018}%
  \BibitemOpen
  \bibfield  {author} {\bibinfo {author} {\bibfnamefont {K.~V.}\ \bibnamefont
  {Lezhnin}}, \bibinfo {author} {\bibfnamefont {P.~V.}\ \bibnamefont
  {Sasorov}}, \bibinfo {author} {\bibfnamefont {G.}~\bibnamefont {Korn}}, \
  and\ \bibinfo {author} {\bibfnamefont {S.~V.}\ \bibnamefont {Bulanov}},\
  }\href@noop {} {\bibfield  {journal} {\bibinfo  {journal} {Phys. Plasmas}\
  }\textbf {\bibinfo {volume} {25}},\ p.\ \bibinfo {pages} {123105} (\bibinfo
  {year} {2018})}\BibitemShut {NoStop}%
\bibitem [{\citenamefont {Wang}\ \emph {et~al.}(2020)\citenamefont {Wang} \emph
  {et~al.}}]{Arefiev-2020}%
  \BibitemOpen
  \bibfield  {author} {\bibinfo {author} {\bibfnamefont {T.}~\bibnamefont
  {Wang}} \emph {et~al.},\ }\href@noop {} {\bibfield  {journal} {\bibinfo
  {journal} {Phys. Rev. Applied}\ }\textbf {\bibinfo {volume} {13}},\ p.\
  \bibinfo {pages} {054024} (\bibinfo {year} {2020})}\BibitemShut {NoStop}%
\bibitem [{\citenamefont {Deng}\ \emph {et~al.}(2002)\citenamefont {Deng} \emph
  {et~al.}}]{osiris-paper}%
  \BibitemOpen
  \bibfield  {author} {\bibinfo {author} {\bibfnamefont {S.}~\bibnamefont
  {Deng}} \emph {et~al.},\ }\href@noop {} {\bibfield  {journal} {\bibinfo
  {journal} {AIP Conference Proceedings}\ }\textbf {\bibinfo {volume} {647}},\
  p.\ \bibinfo {pages} {219} (\bibinfo {year} {2002})}\BibitemShut {NoStop}%
\bibitem [{\citenamefont {Mackenroth}\ and\ \citenamefont
  {Piazza}(2011)}]{DiPiazza-2011}%
  \BibitemOpen
  \bibfield  {author} {\bibinfo {author} {\bibfnamefont {F.}~\bibnamefont
  {Mackenroth}}\ and\ \bibinfo {author} {\bibfnamefont {A.~D.}\ \bibnamefont
  {Piazza}},\ }\href@noop {} {\bibfield  {journal} {\bibinfo  {journal} {Phys.
  Rev. A}\ }\textbf {\bibinfo {volume} {83}},\ p.\ \bibinfo {pages} {032106}
  (\bibinfo {year} {2011})}\BibitemShut {NoStop}%
\bibitem [{\citenamefont {Berestetskii}, \citenamefont {Lifshitz},\ and\
  \citenamefont {Pitaevskii}(1980)}]{LL-QED}%
  \BibitemOpen
  \bibfield  {author} {\bibinfo {author} {\bibfnamefont {V.~B.}\ \bibnamefont
  {Berestetskii}}, \bibinfo {author} {\bibfnamefont {E.~M.}\ \bibnamefont
  {Lifshitz}}, \ and\ \bibinfo {author} {\bibfnamefont {L.~P.}\ \bibnamefont
  {Pitaevskii}},\ }\href@noop {} {\emph {\bibinfo {title} {Quantum
  Electrodynamics}}}\ (\bibinfo  {publisher} {Pergamon Press},\ \bibinfo
  {address} {London},\ \bibinfo {year} {1980})\BibitemShut {NoStop}%
\bibitem [{\citenamefont {Press}\ \emph {et~al.}(2007)\citenamefont {Press},
  \citenamefont {Teukolsky}, \citenamefont {Vetterling},\ and\ \citenamefont
  {Flannery}}]{NumericalRecipes}%
  \BibitemOpen
  \bibfield  {author} {\bibinfo {author} {\bibfnamefont {W.~H.}\ \bibnamefont
  {Press}}, \bibinfo {author} {\bibfnamefont {S.~A.}\ \bibnamefont
  {Teukolsky}}, \bibinfo {author} {\bibfnamefont {W.~T.}\ \bibnamefont
  {Vetterling}}, \ and\ \bibinfo {author} {\bibfnamefont {B.~P.}\ \bibnamefont
  {Flannery}},\ }\href@noop {} {\emph {\bibinfo {title} {Numerical Recipes 3rd
  Edition: The Art of Scientific Computing}}}\ (\bibinfo  {publisher}
  {Cambridge University Press},\ \bibinfo {address} {New York, NY},\ \bibinfo
  {year} {2007})\BibitemShut {NoStop}%
\bibitem [{\citenamefont {Stark}, \citenamefont {Toncian},\ and\ \citenamefont
  {Arefiev}(2016)}]{Arefiev-2016}%
  \BibitemOpen
  \bibfield  {author} {\bibinfo {author} {\bibfnamefont {D.~J.}\ \bibnamefont
  {Stark}}, \bibinfo {author} {\bibfnamefont {T.}~\bibnamefont {Toncian}}, \
  and\ \bibinfo {author} {\bibfnamefont {A.~V.}\ \bibnamefont {Arefiev}},\
  }\href@noop {} {\bibfield  {journal} {\bibinfo  {journal} {Phys. Rev. Lett.}\
  }\textbf {\bibinfo {volume} {116}},\ p.\ \bibinfo {pages} {185003} (\bibinfo
  {year} {2016})}\BibitemShut {NoStop}%
\end{thebibliography}%

\end{document}